\documentstyle[11pt,newpasp,twoside,epsf]{article}
\def\edcomment#1{\iffalse\marginpar{\raggedright\sl#1\/}\else\relax\fi}
\reversemarginpar

\begin{document}
\title{Structural properties of spiral galaxies with and without an AGN: morphology and kinematics}
 \author{Isabel M\'arquez}
\affil{Instituto de Astrof\'{\i}sica de Andaluc\'{\i}a (CSIC), Apdo. 3004, 18080 Granada (Spain)}
\author{DEGAS Consortium}

\begin{abstract}
The analysis of a sample of 30 isolated spiral galaxies with and
without an AGN results in active and control galaxies having similar
bulge and disk properties (since both samples are selected to match in
morphological and bar classification, they have similar bar
percentages). Secondary bars seem also to be
indistinguishable. The central kinematics (both stellar and gas) are
also studied in order to put some constraints on the processes
required to fuel AGNs. Together with previous results, this points to
AGNs being an episode of the life of galaxies, providing that the fuel is
available and the accumulation mechanisms are at work.
\end{abstract}

\section{Introduction}
AGN activity in galaxies is a highly energetic phenomenon
that takes place in a very small spatial extension located at the very
center of some galaxies, and that results to be more frequent in the
Early Universe. The connection between AGN activity and gravitational
interaction was already claimed in the 70's and is still nowadays a
matter of debate.  In any case, galaxies with AGNs are found in
interacting systems, from nearby Seyferts to quasars. Gravitational
interaction is also assumed to increase the percentage of systems
hosting starburst activity. AGN and violent star formation (SF) activity very
often appear simultaneously, but the physical connection between both
phenomena still have to be elucidated.

\section{AGNs and bars}

Interestingly, the morphology of the host galaxies seems to play a
role in determining what galaxies are susceptible of harboring an AGN,
in the sense that AGNs more probably reside in early type spirals
(Balick \& Heckman 1982; M\'arquez \& Moles 1994; Moles, M\'arquez \&
P\'erez 1995; Ho et al. 1997; Knapen et al. 2000). This is specially
important for active galaxies that can be considered as isolated
objects, and therefore the effect of the interaction cannot be invoked
to explain the presence of AGN activity; the presence of a bar appears
as a natural explanation, since it can produce an effective loss of
angular momentum and therefore be the first step in feeding the
central parts with fresh gas coming from the disk (Simkim, Su \&
Schwarz 1980). However, the general result is that the bar percentage
is similar for active and non-active galaxies (Moles et al. 1995;
McLeod \&Rieke 1995; Ho et al. 1997; Mulchaey \& Regan 1997). Hunt \&
Malkan (1999) reach the same conclusion, and also claim that the
incidence of outer and inner rings seem to be different for the
various activity classes (a result that they explain as an
evolutionary sequence from LINERS, which have more inner rings to
Seyferts which have more outer rings). Only Knapen et al. (2000)
obtain a slightly higher fraction of barred galaxies among
Seyferts. Note that when only isolated galaxies are considered, barred
and unbarred galaxies are equally found among active and non-active
galaxies (see M\'arquez et al. 2000).

An additional step is required to transport the material eventually
feeding the AGN close enough to the center. With this respect, nested
bars supply such a mechanism, nuclear bars
producing the same effect as large scale ones, but at smaller
distances to the center (Norman \& Silk 1983; Shlosman et al. 1989;
Wozniak et al. 1995; Friedli et al. 1996). Following this reasoning,
nuclear bars have been looked for, but only found for much fewer cases
than expected if they have to be directly connected with the presence
of nuclear activity (Regan \& Mulchaey 1999; Martini \& Pogee 1999;
M\'arquez et al. 1999, 2000); even if the percentage of nuclear bars
could increase when the identification of a nuclear bar is properly
done, other mechanisms that can also explain the funneling to the
center seem to occur, as nuclear spirals or nuclear warps.

\section{The DEGAS' contribution}

The DEGAS (Dynamics and nuclear Engine of GAlaxies of Spiral
type) project is aimed at studying the connection between the galactic
structure of spiral galaxies and their AGN activity. We have only
selected isolated objects in order to avoid any bias from the inclusion
of interacting objects, the central properties of which could have been
modified by interactions. The active galaxies have been chosen with
the following criteria: (a)~Seyfert 1 or 2 from the V\'eron-Cetty \&
V\'eron (1993) catalogue; (b)~with morphological information in the
RC3 Catalogue; (c)~isolated, in the sense of not having a companion
within 0.4 Mpc (H$_0$=75 km/s/Mpc) and cz$<$500 km/s, or companions
catalogued by Nilson without known redshift; (d)~nearby, cz$<$6000
km/s; and (e)~intermediate inclination (30 to 65$^\circ$). The non
active sample galaxies have been selected among spirals verifying the
same conditions (b), (c), (d) and (e), and with morphologies (given by
the complete de Vaucouleurs coding, not just the Hubble type, so they
also match in bar types) similar to those of the active spirals. We
already stressed that these non active galaxies are well suited to be
used as a control sample. Based on previous work (Moles et al. 1995),
we search for detailed morphological and kinematical differences
between active and non active galaxies of similar global
morphology. We particularly pay attention to those that could
facilitate the transport of gas towards the very central regions and
the nucleus. For this purpose, we are obtaining optical and
near-infrared images and long slit spectroscopy with the best possible
spatial resolution.

\subsection{Near-infrared morphology}
Infrared imaging is particularly important because it allows to trace
the old stellar population, and to separate the various components
(the bulge, disk, bar(s) and spiral arms) with the smallest
contribution of the active nucleus and less contamination by dust
absorption.  We have analysed the morphological and photometric
properties of a sample of isolated spirals with (18) and without (11)
an active nucleus, based on J and K' imaging (M\'arquez et al. 1999, 2000).

The mean resolution of our images is about 1 arcsecond, corresponding
to a physical resolution between 100 and 300 parsecs for the closest
and the more distant galaxy respectively. This resolution implies that
we are able to map the region where the dynamical resonances are
expected to occur (see for instance P\'erez et al. 2000) and is
therefore well suited for our purposes.

We have found that four (one) active (control) galaxies previously
classified as non-barred turn out to have bars when observed in the
near-infrared. One of these four galaxies (UGC 1395) also harbours a
secondary bar. For 15 (9 active, 6 control) out of 24 (14 active, 10
control) of the optically classified barred galaxies (SB or SX) we
find that a secondary bar (or a disk, a lense or an elongated ring) is
present. Our study shows that both sets of
galaxies are similar in their global properties: they define the same
Kormendy relation, their disk components share the same properties,
the bulge and disk scale lengths are correlated in a similar way, bar
strengths and lengths are similar for primary bars. Our results
therefore indicate that hosts of isolated Seyfert galaxies have bulge
and disk properties comparable to those of isolated non active
spirals. Central colors (the innermost 200 pc) of active galaxies are
redder than the centers of non active spirals, most probably due to
AGN light being re-emitted by the hot dust and/or due to circumnuclear
SF, through the contribution of giants/supergiants.

Central to our analysis is the study of the possible connection
between bars and similar non axisymmetric structures with the nuclear
fuelling. We notice that only one of the Seyfert galaxies in our
sample, namely ESO 139-12, does not present a primary bar. But bars
are equally present in active and control objects. The same applies to
secondary bars. Not all the active galaxies we have observed have
them, and some control galaxies also present such central
structures. Secondary central elongations (associated with secondary
bars, lenses, rings or disks) may be somewhat different, but this
result should be confirmed with larger samples.  We note that
numerical models indicate that such secondary bars are not strictly
necessary to feed the central engine when a primary bar is
present. Our results show that down to scales of 100-300 pc, there are
no evident differences between active and non active spiral galaxies.
The details of the data description and analysis are given in
M\'arquez et al. (1999, 2000).

\subsection{Gas and stellar kinematics: the case of NGC~6951}
Nuclear activity needs to be fueled with a supply of gas, the
reservoir of which can be provided by the disk of spiral galaxies. The
existence of gas to fuel the circumnuclear activity is necessary but
not sufficient (e.g., Moles et al. 1995; Mulchaey \& Regan 1997). The
right dynamical and physical conditions must exist for this gas to be
used effectively in either infalling on to the nucleus itself and
feeding the AGN or nuclear starburst, or collapsing by self-gravity in
the circumnuclear region in the form of intense SF. To understand what
are the conditions and mechanisms for the onset of nuclear activity in
spiral galaxies, a detailed characterisation of morphological and
kinematical components in galaxies of different morphology and
activity level is needed.  We have already proceeded with such an
analysis for the active SAB(rs)bc galaxy NGC 6951, by means of broad
band B'IJK images and high resolution high dispersion longslit
spectroscopy, together with archival HST WFPC2 V and NICMOS2 J and H
images (P\'erez et al. 2000). We found that:

{\bf a)} There is little ongoing SF inside the bar dominated
region of the galaxy, except for the circumnuclear ring at 5\arcsec
radius. This SF occurs in two modes, in bursts and
continuously, along the ring and inwards towards the nucleus. The
equivalent width of the CaII triplet absorption lines show that in the
metal rich central region of this galaxy, within 5\arcsec radius, the
continuum light is dominated by a population of red supergiant stars,
while outside the circumnuclear ring the stellar population is that of
giants.

\begin{figure}
\plotfiddle{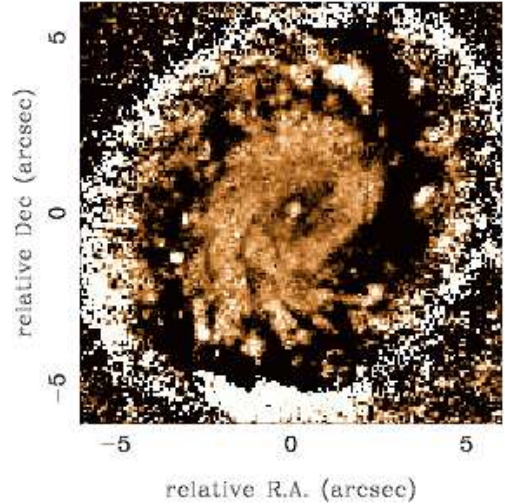}{6cm}{0}{60}{60}{-190}{-150}
\caption{J-H colour image of the
circumnuclear region of NGC 6951 (HST+NICMOS, spatial scale =
0.075\arcsec/px) where the nuclear spiral is clearly seen.}
\end{figure}


\begin{figure}
\plotfiddle{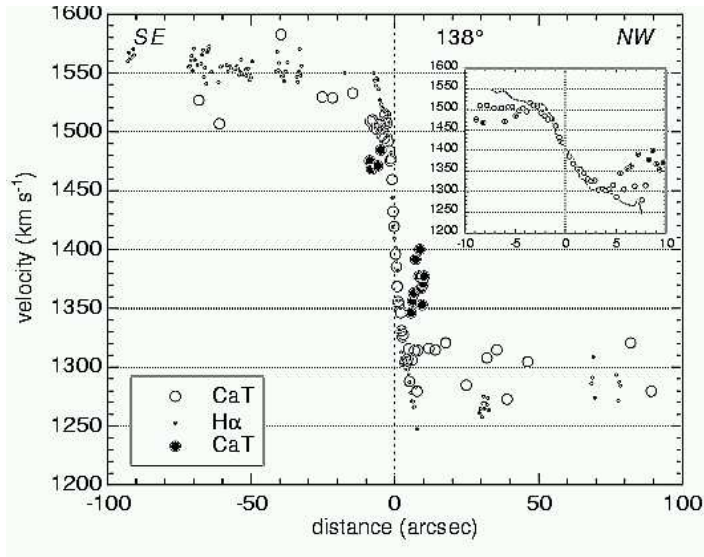}{7cm}{0}{50}{50}{-120}{0}
\caption{Stellar (circles) and ionized gas (dots) velocity
curve along the major axis of NGC 6951.  
In the central region the CaT absorption lines are resolved
in two components (filled and open circles). }
\end{figure}

{\bf b)} We suggest that the gaseous and stellar kinematics along the
three slit position angles can be interpreted as the existence of a
hierachy of disks within disks, with dynamics decoupled at the two
ILRs, that we find to be located at 180 pc and at 1100 pc. 
This would be supported by the nuclear spiral
structure seen in the high resolution HST images (see Fig. 1).
Outside the iILR the stellar CaT velocity
profile can be partly resolved into two different components that seem
to be associated with the bar and a disk (see Fig. 2).

{\bf c)} We discuss the possibility that the kinematic component
inside the iILR could be due to a nuclear disk, as in the disk within
disk scenario suggested above, or to a nuclear outflow produced by the
combined effects of SN and SN remnants.  Several clues indicate that
this is a dynamically old system: (i) there is little SF ongoing
inside the bar dominated part of the galaxy (except for the
circumnuclear ring), (ii) the relative amount of molecular to total
mass within the inner 6 arcsec radius is very large $\approx$ 25\%,
and (iii) the geometry of the circumnuclear ring leading at a position
angle greater than 90\deg~ from the stellar bar. It is thus possible
that a nuclear bar has existed in NGC 6951 that drove the gas towards
the nucleus, as in the bars within bars scenario, but that this bar
has already dissolved due to the accumulation of the gas within the
circumnuclear region (P\'erez et al. 2000).

A detailed kinematical and morphological analysis for a number of
galaxies in our sample is in progress. Intermediate resolution data
for the gas kinematics have already been obtained for the great
majority of the galaxies. High spatial and spectral resolution
spectroscopy for the gas and stars have been obtained for about ten
galaxies and will be analysed as for NGC 6951.

\section{Conclusions and further considerations}
We have shown that isolated spiral galaxies with or without an AGN are
equivalent in the sense of hosting similar large scale components
(bulges, disks, bars). To analyse the processes taking place in the
central regions, both detailed morphology and high resolution
spectroscopy of gas and stars are needed, together with detailed
models for the interpretation, as it is the case for example for the double
stellar component in NGC 6951.

Bars are dynamically active, so it is relevant to consider that
different time scales are involved in the processes of bar creation
and evolution, SF and AGN activity, and hence the observable
connection between bars and AGNs may be more complex. With respect to
the availability of fuel, it has been found that CO is more
concentrated in barred galaxies, with no difference for active and
non-active ones (Sakamoto 1999). Similar rotation curves (Sofue \&
Rubin 2001) and environments have also been reported for active and
non-active galaxies. The masses of the central compact objects (black
holes) seem to be comparable (Ferrarese \& Merritt 2000; McLure \&
Dunlop 2000; Laor 2001), with M$_{bh}$/M$_{bulge}$ $\approx$ 0.005\%,
with a maximum of $\approx$ 0.025\% (we stress that numerical
simulations show that the bar is destroyed when the central mass is
about 1-10\% the total mass -- see Combes 2001--, so this could
explain the limit in BH masses).  Therefore, if galaxies hosting AGN
have no properties differing from those with of galaxies without an
AGN, it could be that the AGN is an episode of the life of a galaxy,
providing that the fuel is available and the accumulation mechanisms
are at work. Such accumulations would also help to trigger SF
processes (competitive or not) that seem to coexist with AGN activity.

\end{document}